\documentclass{icrc2009}

\usepackage{graphicx}   
\usepackage{fixltx2e}

\newcommand{\shorttitle}[1]%
{\markboth{Proceedings of the 31\MakeLowercase{$^{st}$} ICRC, {\L}\'{o}d\'{z} 2009}{#1} }
\newcommand{\etal}{\MakeLowercase{\textit{et al. }}} 

\begin{document}
\title{The Tunka-133 EAS Cherenkov array - status, first results and plans}

\author{\IEEEauthorblockN{\\N.M. Budnev\IEEEauthorrefmark{2},
			  D. Besson\IEEEauthorrefmark{7},
                          O.A. Chvalaev\IEEEauthorrefmark{2},
                           O.A. Gress\IEEEauthorrefmark{2},
			   N.N. Kalmykov\IEEEauthorrefmark{1},
			   A.A. Kochanov\IEEEauthorrefmark{2},
			   \\
			   A.V. Korobchenko\IEEEauthorrefmark{2},
			   E.E. Korosteleva\IEEEauthorrefmark{1},
			   V.A. Kozhin\IEEEauthorrefmark{1},
			   L.A. Kuzmichev\IEEEauthorrefmark{1},
			   B.K. Lubsandorzhiev\IEEEauthorrefmark{3},
			   \\
			   R.R. Mirgazov\IEEEauthorrefmark{2},
			   G. Navarra\IEEEauthorrefmark{4},
			   M.I. Panasyuk\IEEEauthorrefmark{1},
			   L.V. Pankov\IEEEauthorrefmark{2},
			   V.V. Prosin\IEEEauthorrefmark{1},
			   V.S. Ptuskin\IEEEauthorrefmark{5},
			   \\
			   Yu.A. Semeney\IEEEauthorrefmark{2},
			   B.A. Shaibonov(junior)\IEEEauthorrefmark{3},
			   A.V. Skurikhin\IEEEauthorrefmark{1},
			   J. Snyder\IEEEauthorrefmark{7},
			   C. Spiering\IEEEauthorrefmark{6},
			   M. Stockham\IEEEauthorrefmark{7},
			   \\
			   R. Wischnewski\IEEEauthorrefmark{6},
			   I.V. Yashin\IEEEauthorrefmark{1},
			   A.V. Zablotsky\IEEEauthorrefmark{1},
                           A.V. Zagorodnikov\IEEEauthorrefmark{2}}
                            \\
\IEEEauthorblockA{\IEEEauthorrefmark{1}Skobeltsyn Institute of Nuclear Physics
MSU, Moscow, Russia}
\IEEEauthorblockA{\IEEEauthorrefmark{2}Institute of Applied Physics ISU, Irkutsk,
Russia}
\IEEEauthorblockA{\IEEEauthorrefmark{3}Institute for Nuclear Research of RAN,
Moscow, Russia} 
\IEEEauthorblockA{\IEEEauthorrefmark{4}Dipartimento di Fisica Generale
Universiteta di Torino and INFN, Torino, Italy}
\IEEEauthorblockA{\IEEEauthorrefmark{5}IZMIRAN, Troitsk, Moscow Region, Russia}
\IEEEauthorblockA{\IEEEauthorrefmark{6}DESY, Zeuthen, Germany }
\IEEEauthorblockA{\IEEEauthorrefmark{7}Department of Physics and Astronomy,
University of Kansas,USA}}

\shorttitle{L.Kuzmichev\etal The Tunka-133 EAS Cherenkov array}
\maketitle

\begin{abstract}
 
The new EAS Cherenkov array Tunka-133 with about 1 km$^2$ geometric acceptance
area, is installed in the Tunka Valley (50 km from  Lake Baikal). The array
will permit a detailed study of cosmic ray energy spectrum and mass composition
in the energy range of 10$^{15}$ - 10$^{18}$ eV with a uniform method. The array
consists of 19 clusters, each composed of 7 optical detectors with 20 cm PMTs.
Since November 2008, the first 12 clusters are in operation, commissioning of
the whole array is planned for September 2009\footnote{At the time of submission of this paper to the
electornic arXiv(February 2010) the comleted Tunka-133 array is already taking data}. We describe the array
construction and DAQ, preliminary results and plans for the future development:
deployment of radio-antennas and muon detectors network. 

\end{abstract}

 
\section{Introduction}

The elaborate study of primary mass composition in the energy range $10^{15} -
10^{18}$ eV is of crucial importance for the understanding of the origin and
propagation of cosmic rays in the Galaxy. Since 2006 the work on creation of the
EAS Cherenkov array Tunka-133 with a geometric area of 1 km$^2$
(\cite{Tunka1}, \cite{Tunka2}) is carried out in the Tunka Valley. Such an array will
allow the investigation of cosmic rays in the energy range from 10$^{15}$ to
10$^{18}$ eV by a uniform method. This energy range includes the knee in the
energy spectrum at 3$\cdot$10$^{15}$ eV, and other features of the spectrum
probably connected with the transition from galactic to extragalactic 
cosmic rays. For one year operation (400 hours) the array will register more
than 300 events with energy above 10$^{17}$eV and the core position inside the array
geometry. The use of algorithms of reconstruction of EAS parameters developed 
for the Tunka-25 array \cite{Tunka25} will provide an accuracy of the measurement
of EAS maximum depth $X_{max}$ $\sim$ 25 g$\cdot$ cm$^{-2}$. The Tunka-133 array
has been detecting Cherenkov light from EAS over the last two years, with a
steadily increasing number of detectors (7 detectors -- 2006, 28 detectors --
2007 and 84 detectors -- 2008) and array effective area. The final array of
133 detectors will be completed by autumn 2009, and the internal effective area will
reach 1 km$^2$.

\section{Tunka-133 array -- short description}
 
The whole Tunka-133 array will consist of 133 optical detectors on the basis
of PMT EMI 9350 \mbox{(20 cm} photocathode diameter). Detectors are grouped into
19 clusters with seven detectors in each one -- six hexagonally arranged
detectors and one in the center. The distance between the detectors is 85 m.\\ 
An optical detector consists of a metallic cylinder of 50 cm diameter, containing
a PMT. The container window is directed to zenith and covered with plexiglas
heated against hoar-frost and dew. The detector is equipped with a remotely
controlled lid protecting the PMT from sunlight and precipitation. The detector
efficiency reduces smoothly to about 80\% of the vertical one at a zenith angle of
40$^{\circ}$ and to 50\% at 50$^{\circ}$. The zenith angle distribution of EAS
with energy more than 7 PeV is presented in Fig.\,1. It indicates the 
necessary correction
due to the dependence of the efficiency on the angle 
which is applied during the EAS energy reconstruction. So the new
detector allows analyzing events with zenith angles up to 50$^{\circ}$.\\ 
In addition to the PMT with its high voltage supply and preamplifiers, 
the container houses a light
emitting diode for calibration \cite{LED} and a detector controller. 
The controller is connected to the cluster electronics via
twisted pair cable by the RS-485 protocol. To provide the necessary dynamic range of
10$^4$, two analog signals, one from the anode and the other from 
the dynode are read out.
The ratio of amplitudes of these signals is about 30. Both signals are
transmitted to the cluster electronics hut via  100 m RG-58 coaxial cables.\\  
The cluster electronics \cite{FADC} consists of a cluster controller, 4 blocks
of four-channel FADCs, an adapter block connected with 7 detector controllers,
and the separate temperature controller.
The cluster controller consists of an optical transceiver, a synchronization 
module, a local time clock and a trigger module. The optical transceiver operating
at 1000 MHz is responsible for data transmission and formation of a 
100 MHz synchronization
signal for cluster clocks. The cluster trigger (the local trigger) is
formed by the coincidence of at least three pulses from detectors exceeding the
threshold within a time window of 0.5$\mu$s. The time mark of the local trigger
is fixed by the cluster clock. The accuracy of the time synchronization between
different clusters is about 10\,ns. The FADC boards are designed on the basis of
12-bit 200\,MHz ADCs and XILINX microchip FPGAs. 
All the electronic blocks (except the temperature controller) are implemented in
VME standard. Each cluster electronics is connected to the DAQ center with a
multi-wire cable consisting of four copper wires and four optical fibers.
     
 \begin{figure}[!t]
  \centering
  \includegraphics[width=2.5in]{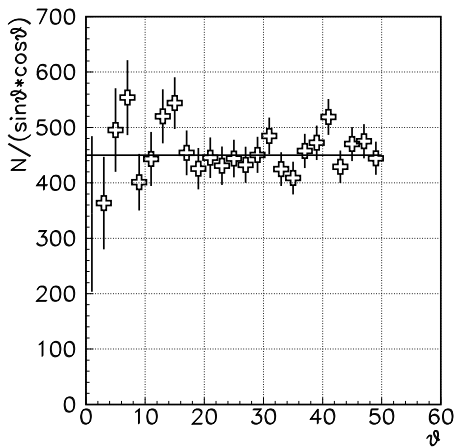}
  \caption{Zenith angle distribution of events with energy $>$ 7$\cdot$10$^{15}$
  eV} 
  \label{simp_fig}
 \end{figure}
 
\section{Preliminary data analysis}

The first cluster (7 detectors) has acquired data for about 75 hours from
November 2006 to January 2007. Four
clusters were operating over the
next winter, from November 2007 to April 2008. Data
have been  
recorded over 270 hours during clean moonless nights. The average trigger rate
was about 0.3 Hz, the number of the registered events was about 300\,000. The low
energy part of the spectrum is shown in Fig.\,2 together with the Tunka-25
spectrum. The energy of 100\% efficiency is about 2$\cdot$10$^{15}$ eV. It will
be possible to decrease the energy threshold of the array by selecting showers with
their core position located inside of the cluster geometric area.\\ 
Since November 2008, 12 clusters were in operation. Data from this winter season
are still under analysis and the preliminary results would  be presented at the
conference. 

 \begin{figure}[!t]
  \centering
  \includegraphics[width=2.5in]{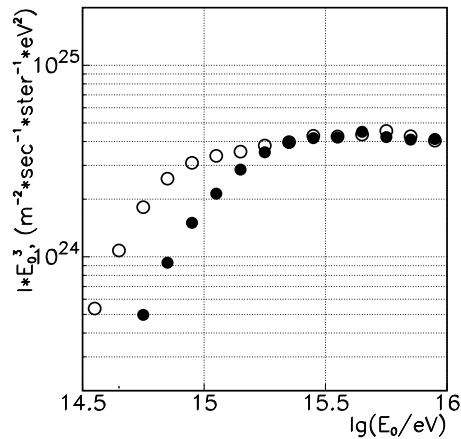}
  \caption{The energy spectrum from Tunka-25 (open circles) and Tunka-133
(filled circles) in the threshold region. } 
  \label{simp_fig}
 \end{figure}
 
The program of calibration and reconstruction of EAS parameters consists of
three main blocks of codes.

1. The first block analyzes the primary data records for each Cherenkov light
detector and derives three main parameters of the pulse: front delay at a level
0.25 of the maximum amplitude ($t_i$), pulse area ($S_i$) and full width at
half-maximum $FWHM_i$.

2. The second block of codes unites the data of different clusters and provides
the relative time and amplitude calibration. Data from various clusters are
merged to one event, if the time difference for cluster triggers is less than 2
$\mu s$. 

The time and amplitude calibration and conversion of $S_i$ to light density
($Q_i$) is provided in the  way described in \cite{Tunka25}.

3. The third block of programs reconstructs the EAS core location, the primary
energy and the depth of the shower maximum. 

 \begin{figure}[!t]
   \centering
   \includegraphics[width=2.5in]{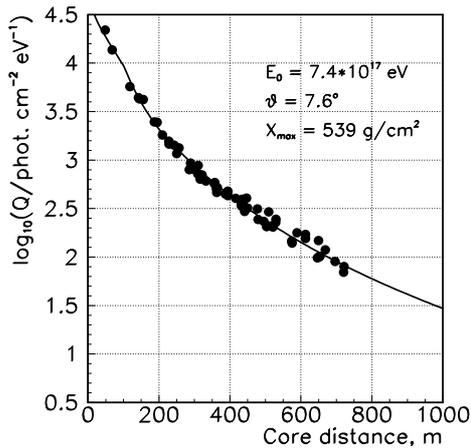}
   \caption{ Lateral distribution resulting from fitting the measured light fluxes
(points)}
\label{simp_fig}
\end{figure}
he first method of EAS core location reconstruction is fitting the $Q_i$ (light
density at detector $i$) by the lateral distribution function (LDF) with
variable 
parameter of steepness ($P$) and photon flux density at a core distance 175 m
($Q_{175}$).  This function -- in the form used for Tunka-25 -- was first
presented in 
\cite{LDF}, and it was extended to include large distances in \cite{cris08}.
In addition to the traditional method decribed above, a new method of EAS core
reconstruction using Cherenkov light pulse $FWHM$ has been developed and
implemented in the code. To fit the experimental $FWHM_i$, the simulated
width-distance function (WDF) is used. The analytic form of 
the WDF has been derived from CORSIKA simulated light pulses taking into account
the apparatus distortion of their waveforms. 
In more detail this method of EAS core reconstruction and conversion of
$FWHM(400)$ to $X_{max}$ is discussed in \cite{FWHM}, presented at this
conference. The EAS maximum depth $X_{max}$ will be reconstructed for each event
by two independent methods from LDF steepness $P$ and the parameter $FWHM(400)$.
As an example, the LDF and WDF for one event are shown in Figs. 3 and 4 
respectively.\\ 
It seems that the absence of FWHM random fluctuations and a more simple expression
for WDF instead of LDF will allow us applying the new method of EAS core
reconstruction not only inside the geometrical area of the array, but also 
outside (up to a certain distance). It will be possible to check the accuracy of this method up to
1 km core distance by using events with their core (reconstructed by the LDF method)
located near the array boundary, using for the WDF method only 
those parts of the detector which are far away 
from the core. To check this method up to 1.5 km, the installation of
one or two additional clusters at 1 km distance from the center of the array is
planned. The possibility to reconstruct events up to 1.5 km from the array
center will increase the array effective area for events with energy more than 
$5\cdot 10^{17}$ eV by a factor of 5--10 relative to the geometrical area 
covered by the detectors.

 \begin{figure}[!t]
   \centering
   \includegraphics[width=2.5in]{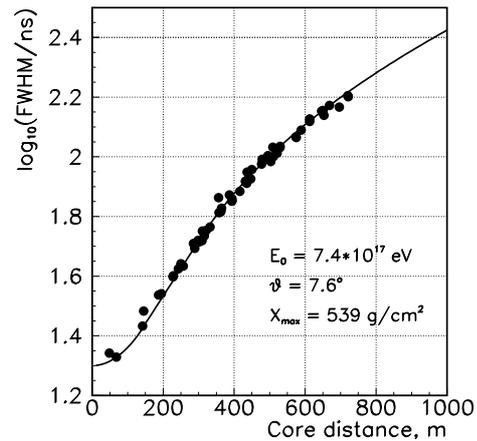}
  \caption{Dependence of the width of the Cherenkov light pulse on the core distance. Points --
    experiment, line -- analytic form of WDF }
    \label{simp_fig}
\end{figure}


\section{Plan for Tunka-133 upgrading}

To increase the effective up-time of the array, 
we plan to upgrade the Cherenkov
array with additional detectors which may register EAS during the day time.  

\subsection{Registration of radio signal from EAS}

The registration of the radio flux from EAS is -- like the registration of Cherenkov
light -- a calorimetric method which may give information about the energy of
primary particles and the depth of the shower maximum with good accuracy. To
cross-check this method, we plan to install several radio antennas for common
operation with the Cherenkov array. The first 2 high-gain log-periodic antennas were
installed in July 2008 (Fig.\,5). The antennas are located at 10 m distance
from the 
cluster electronics box and connected with amplifiers (52 dB and 60 dB
respectively) via 20 m LMR-400 coaxial cable. Signals from amplifiers, placed
inside the cluster electronics box, are sent to two free channels of \mbox{Tunka-133}
FADC units. These channels (in standard operation mode) are not included in
cluster trigger formation, and so the data from the
antennas are read out only if the cluster
trigger formed by the optical detectors occurs. 

In addition to Tunka-133 data
storage, data from a third test antenna are also captured at 10-minute intervals
on a laptop (using LabView) to allow constant monitoring of changes of the RF
environment. The Tunka site shows practically no background in the FM region (88 -
108 MHz) and, with ecxeption of a 1 - 2 MHz interval around 80 MHz, a quite low
background in the region $>$40 MHz (Fig.\,6), which is a range typical of radio air
shower detection experiments. MC simulation showed that the signal from 5$\cdot$10$^{16}$
eV EAS at a distance \mbox{$\sim$ 100 m} from the EAS core should be clearly visible
above background. Experimental data from the winter season is currently
under analysis, and preliminary results will be presented at the conference.
Additional 4 antennas are planned to be installed in summer 2009.

\subsection{Registration of muons}

The deployment of scintillation counters within the Tunka array would provide a
cross-calibration of different methods of EAS parameter measurement since all
components will be recorded simultaneously. Measuring of Cherenkov light
provides the energy of the primary particle and the depth $X_{max}$ of the EAS maximum.
The scintillation counters, buried under a layer of 1.5 - 2 m of ground, would
allow an 
estimate of the number of muons in the same events, resulting in a more precise
mass composition in the energy range of \mbox{10$^{17}$ - 10$^{18}$ eV}. These methods
were suggested by G.~Kristiansen and collaborators already in 1981 \cite{KGB}.
    
\begin{figure}[!t]
\centering
\includegraphics[width=2.5in]{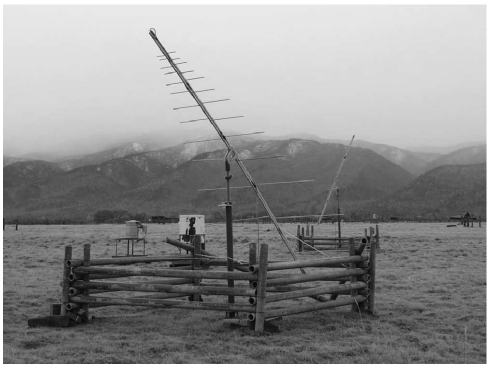}
\caption{View of the radio antennas}
\label{simp_fig}
\end{figure}

The new simulation performed with the AIRES code \cite{muon_sim} showed that
the measurement of the muon number with an accuracy of ($\sim$ 5 - 10\%),
together with energy and $X_{max}$ with the accuracy achieved at Tunka-133, increases
the capability to distinguish Fe from CNO showers. To reach
a sufficient accuracy for measurement of the muon number, 
the spacing of the muon detectors should not exceed
120-150 m. It was shown \cite{muon_sim} that by using 20 muon detectors with
10 m$^2$ area placed inside the geometric area of Tunka-133 it is possible to
reach an accuracy of about 10\% in the total muon number at 10$^{17}$ eV.
  
\section{Other research items}
\subsection{Cosmic rays and atmospheric electricity processes }

During the summer season we continue the study of the variation of secondary
cosmic 
rays (electrons and muons) correlated with the variation of the electric field
at the ground level and the space-time correlation between lightnings and high
energy EAS. 
In these experiments a water Cherenkov light detector with \mbox{10 m$^2$} area
and several radio antennas are used. 

\subsection{Robotic Optical telescope}

For the purpose of continuous sky monitoring, the optical telescope MASTER
(Mobil Astronomy System of TElescope Robots) \cite{MASTER} will be installed
during this summer at Tunka. The main scientific topic of the system is the
search for optical radiation from GRBs. For the operation of Tunka-133 this
system will give very important information about atmospheric conditions
(variation of atmosphere transparency and existence of clouds).

\begin{figure}[!t]
\centering
\includegraphics[width=2.5in]{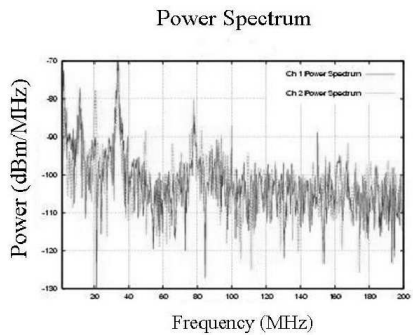}
\caption{Raw spectral power as a function of frequency for two
	    antennas. The thermal floor at 300 K is expected to 
	    be - 114  dBm/MHz.}
\label{simp_fig}
\end{figure}

\section{Conclusion }

The deployment of the Tunka-133 array is approaching its completion.  More than
75\% 
of detectors are installed and operating since the last winter season.
Commissioning of the complete array is planned in autumn 2009.\\ 
The first results show that the array parameters are in a good agreement with
the expected ones. Recording of the EAS Cherenkov light pulse waveform for each
detector provides a more reliable measurement of $X_{max}$ and opens the
possibility to reconstruct EAS with their core outside the array.
Works for upgrading Tunka-133 with day-time operating detectors (radio-antennas
and muon scintillation counters) have started. 
  
\section{Acknowledgements}  

The present work is supported by the Russian Ministry of Education and Science,
program "Development of Scientific Potential of High Energy Schools"(projects
2.2.1.1/1483, 2.1.1/1539, 2.2.1/5901), by the Russian Foundation of Basic Research
(grant 07-0200904, 09-02-1000, 09-02-12287) and by the Deutsche
Forschungsgemeinschaft DFG (436 RUS 113/827/0-1)


\begin{thebibliography}{99}
   
\bibitem{Tunka1} N.M. Budnev et al., (Tunka Collaboration), Proc. 29th ICRC,
Pune, India, 8 (2005) 255, arXiv: astro-ph/0511229  
                     
\bibitem{Tunka2}   N.M. Budnev et al., (Tunka Collaboration), Proc. 30th ICRC,
Merida, Yucatan, Mexico, 
  5 (2007) 973, arXiv: 0801.3037
                    
\bibitem{Tunka25}   N.M. Budnev et al., (Tunka Collaboration), Proc. 29th
ICRC,Pune, 
India, 6 (2005) 257, 
    arXiv: astro-ph/0511220	    
	
\bibitem{LED} Lubsandorzhiev B.K. et al.,  Proc. 30th ICRC, Merida, Yucatan,
Mexico, (2007), 
                arXiv:  0709.0458
    	       
\bibitem{FADC}   N.M.Budnev et al., Proc. 10th ICATPP, Italy 2007, 
    World Scientific, 2008, pp 287 - 291, arXiv: 0804.0856  
    	    
\bibitem{LDF}  EAS-TOP Collaboration and Korosteleva E.E., Kuzmichev L.A.,
   Prosin V.V., 
   Proc. 28th ICRC, Tsukuba, Japan, 1 (2003) 89 
   
\bibitem{cris08} N. Budnev et al., (Tunka Collaboration), Nucl. Phys. B (Proc.
Suppl.) 190 (2009) 247 - 252, arXiv: 0902.3156
    
\bibitem{FWHM}  E.E. Korosteleva, L.A. Kuzmichev, V.V. Prosin, This
   proceedings, ID = 492 
       	    
\bibitem{KGB}  V.V. Atrashkevich, N.N. Kalmykov and G.B.Khristiansen,
   JETP letters, 33 (4) (1981) 225-227  
    
\bibitem{muon_sim}  N.N. Kalmykov et al., This proceedings, ID = 1073     
   
\bibitem{MASTER} V.M. Lipunov et al., (2007), arXiv: 0711.0037   
    
\end{thebibliography}
\end{document}